\def\year{2022}\relax
\documentclass[letterpaper]{article} 
\usepackage{aaai22}  
\usepackage{times}  
\usepackage{helvet}  
\usepackage{courier}  
\usepackage{colortbl}
\usepackage[hyphens]{url}  
\usepackage{graphicx} 
\usepackage{natbib}  
\usepackage{caption} 
\DeclareCaptionStyle{ruled}{labelfont=normalfont,labelsep=colon,strut=off} 
\usepackage{algorithm}
\usepackage{algorithmic}

\usepackage{newfloat}
\usepackage{listings}
\floatstyle{ruled}
\newfloat{listing}{tb}{lst}{}
\floatname{listing}{Listing}
\title{A Study of Partisan News Sharing in the Russian invasion of Ukraine\thanks{\textbf{[Accepted by ICWSM 2024]}}}
\author {
    Yiming ZHU\textsuperscript{\rm 12}
    Ehsan-Ul Haq\textsuperscript{\rm 1}
    Gareth Tyson\textsuperscript{\rm 1} 
    Lik-Hang Lee\textsuperscript{\rm 3}
    Yuyang Wang\textsuperscript{\rm 1}
    and Pan Hui\textsuperscript{14}\\
}
\affiliations {
    \textsuperscript{\rm 1} The Hong Kong University of Science and Technology (Guangzhou)\\
    \textsuperscript{\rm 2} The Hong Kong University of Science and Technology\\
     \textsuperscript{\rm 3} The Hong Kong Polytechnic University\\
     \textsuperscript{\rm 4} The University Of Helsinki\\
    \{yzhucd, euhaq\}@connect.ust.hk \quad \{gtyson, panhui\}@ust.hk
}

\usepackage{graphicx}
\usepackage{amsmath}
\usepackage{caption}
\usepackage[dvipsnames]{xcolor}
\usepackage{subcaption}
\usepackage{multirow}
\usepackage{booktabs,threeparttable}
\usepackage{siunitx}
\usepackage{dcolumn}
\usepackage{csquotes}
\usepackage{enumitem}
\usepackage{makecell}
\usepackage{float}
\usepackage{subcaption}
\usepackage{xspace}
\usepackage{amssymb}
\usepackage{bm}
\usepackage[normalem]{ulem}
\useunder{\uline}{\ul}{}
\usepackage{pifont}
\newcommand{\cmark}{\text{\ding{51}}}

\usepackage{bibentry}

\begin{document}

\maketitle

\newcommand\update[1]{{\textcolor{black}{#1}}}
\newcommand\tbd[1]{\textbf{\textcolor{blue}{TBD: #1}}	}
\newcommand\james[1]{\textbf{\textcolor{cyan}{JZ: #1}}	}

\newcommand\ehsan[1]{\textbf{\textcolor{magenta}{EH: #1}}	}
\newcommand\gareth[1]{\textbf{\textcolor{red}{GT: #1}}	}

\newcommand{\one}{({\em i}\/)\xspace}
\newcommand{\two}{({\em ii}\/)\xspace}
\newcommand{\three}{({\em iii}\/)\xspace}
\newcommand{\four}{({\em iv}\/)\xspace}
\newcommand{\five}{({\em v}\/)\xspace}
\newcommand{\six}{({\em vi}\/)\xspace}
\newcommand{\seven}{({\em vii}\/)\xspace}
\newcommand{\eight}{({\em viii}\/)\xspace}
\newcommand{\nine}{({\em ix}\/)\xspace}

\def\eg{\emph{e.g.}\xspace}
\def\etc{\emph{etc.}\xspace}
\def\ie{\emph{i.e.}\xspace}
\def\etal{\emph{et al.}\xspace}
\def\vs{\emph{vs.}\xspace}
\def\cf{\emph{cf.}\xspace}
\newcommand{\pb}[1]{\vspace{0.75ex}\noindent{\bf \em #1}\hspace*{.3em}}
\newcommand{\pbb}[1]{\vspace{0.75ex}\noindent{\bf #1}\hspace*{.3em}}

\begin{abstract}
Since the Russian invasion of Ukraine, a large volume of biased and partisan news has been spread via social media platforms. 
As this may lead to wider societal issues, we argue that understanding how partisan news sharing impacts users' communication is crucial for better governance of online communities. 
In this paper, we perform a measurement study of partisan news sharing. We aim to characterize the role of such sharing in influencing users' communications.
Our analysis covers an eight-month dataset across six Reddit communities related to the Russian invasion. We first perform an analysis of the temporal evolution of partisan news sharing. We confirm that the invasion stimulates discussion in the observed communities, accompanied by an increased volume of partisan news sharing. Next, we characterize users' response to such sharing.
We observe that partisan bias plays a role in narrowing its propagation. More biased media is less likely to be spread across multiple subreddits.
However, we find that partisan news sharing attracts more users to engage in the discussion, by generating more comments. 
We then built a predictive model to identify users likely to spread partisan news.
The prediction is challenging though, 
\update{with 61.57\% accuracy on average. 
Our centrality analysis on the commenting network further indicates that the users who disseminate partisan news possess lower network influence in comparison to those who propagate neutral news.} 
\end{abstract}

\section{Introduction}


The widespread sharing of partisan news on online social networks (OSNs) has become a prominent concern in recent years.  It has been shown that such partisan media can damage public discourse and trigger echo chambers~\cite{haqWeapon2022}, along with several other long-lasting effects on society, \eg creating ideological bias, engineering public views~\cite{solo2017overview}, and spreading hostility~\cite{muddiman2017news}.

Online news sharing is often driven by large-scale events that engender attention from a wide audience~\cite{koutra2015events}. The increased interest of users in the news during a social event makes users more susceptible to potential problems as highlighted before~\cite{pierri2022propaganda}. Such social events offer unique opportunities to characterize users' communication and the resulting engagement based on the news content being shared. For instance, profiling trolls on Twitter to gain insight into their influence on political discourse during the 2016 U.S. election~\cite{badawy2019falls} and tracing the propagation of misinformation in COVID-19 pandemic 2020~\cite{sharma2020covid}.

One recent exemplar
is the Russian invasion of Ukraine in 2022.
On February \(24^{th}\),
Russia officially launched its ``special military operation'' against Ukraine. 
Subsequently, a large amount of biased and partisan news has been shared online~\cite{yaraduainfluence,osmundsen2022information}
, making this an interesting case study.
Recent works on Reddit have shed light on the important role of partisan news sharing in analyzing the propagation of political narratives~\cite{hanley2022happenstance} and  troll accounts~\cite{saeed2022trollmagnifier}. These works both reveal the importance of understanding the evolution of media dissemination, users' responses to partisan news and the behavior of users who spread partisan news. 
By focusing on six Reddit communities, we strive to analyze partisan news sharing from these three perspectives and characterize its role in influencing users' communication. 

In this paper, we perform a study of partisan news sharing on subreddits relevant to the Russian invasion.
Our analysis covers an eight-month dataset across six relevant subreddits. To identify the URLs from partisan news media, we utilize a large-scale list of rated
domains with partisan bias scores proposed by Robertson~\cite{robertson2018auditing}. Each domain on the list is assigned with a partisan score to categorize its partisan leaning. In all, we annotate 77,871 URLs from partisan news and assign a  partisan score to each user based on the articles they share.
Note, we use the word \enquote{submissions} for both posts and comments on Reddit.

Using this data, we first inspect the temporal evolution of partisan news sharing (Section~\ref{sec:Temporal Evolution}). 
We find that the Russian invasion stimulates users' discussion in the observed subreddits, accompanied by a significant surge of submissions sharing partisan news. We then characterize the sharing of partisan news from three perspectives:
sharing across communities (Section~\ref{sec:range}), users' responses to posts and comments containing partisan news (Section~\ref{sec:response}), and the communication characteristics of partisan spreaders (Section~\ref{sec:PCommunicator}). 



We then investigate whether media with higher partisan bias tends to be spread across more subreddits. 
We find that the media's partisan score does correlate with narrowing media sharing: the more partisan a news item is, the less likely for it to be spread across multiple subreddits (Section~\ref{sec:range}).
This leads us to inspect users' responses to the submissions containing partisan news. Specifically, we focus on the volume of comments and votes that are accumulated across these submissions.
Our analysis shows that users' reactions to these submissions are significantly different from those not containing partisan news. Partisan news receives higher user engagement in discussions by generating more comments. Moreover, users are more likely to upvote posts containing partisan news (Section~\ref{sec:response}).


Further, we profile spreaders who have a high partisan score. Our results reveal that highly biased spreaders are more active in generating partisan content. \update{On average, these spreaders submit more partisan news, and tend to include more URLs from partisan news in their submissions. Additionally, comments submitted by highly biased spreaders are more likely to be commented on, while posts submitted by these spreaders seem to receive lower voting scores. To quantify the most important features, we use linear regression to predict partisan spreader. The evaluation shows that the classifier can achieve an average accuracy 61.57\% (Section~\ref{sec:PCommunicator}).}


\update{Finally, we examine the impact of sharing partisan news within the interaction network.
We specifically explore the impact that sharing partisan news has on the users' influence.
To estimate their influence, we use three different centrality metrics on a directed network derived from comment interactions. Our findings indicate that highly biased spreaders have lower centrality. This suggests that sharing media with a high partisan bias is not an effective strategy for increasing spreaders' influence. Moreover, we find a positive correlation between spreaders' PageRank and the received voting scores in posts (Section~\ref{sec:network}).}

\section{Related Work}

\pbb{News Sharing on Russian Invasion.} News sharing plays an important role in influencing online public discourse during large-scale events~\cite{bolsen2018us}. 
In the context of the Russo-Ukrainian conflict, earlier works have depicted the effect of partisan news sharing in causing problems on OSNs, including rumor cascades~\cite{zannettou2019web} and political polarization~\cite{tkachenko2019conflict}. To reduce such an effect, some researchers have characterized the propagation of partisan news~\cite{zannettou2019let, bobichev2017sentiment}, tracing the source of such propagation~\cite{karamshuk2016identifying}. Hanley \etal{}~\cite{hanley2022happenstance} has taken a closer look at partisan news sharing on \textit{r/russia}, a Reddit Russian community. They reveal that, during the Russian invasion, comments in \textit{r/russia} present more leaning to partisan news from Russian state media.

Caprolu \etal{}~\cite{caprolu2022characterizing} have performed an aspect-based sentiment analysis on more than five million tweets related to the Russian invasion. They discover evidence that there are no massive disinformation campaigns in users' news sharing, in contrast to what is suggested by mainstream media. In our work, focusing on Reddit communities related to the Russian invasion, we characterize partisan news sharing from the aspects of media's domain, posts and comments, as well as the behavior of spreaders. Our results provide an understanding of the crucial role of partisan news sharing in impacting users' communication.

\pbb{Partisan News Spreaders.} 
It has been shown that people who spread partisan news play a part in facilitating the propagation of fake news~\cite{shrestha2019online, shu2019studying}, misleading audiences' cognition to events~\cite{allen2022birds} and exacerbating polarization~\cite{lima2018inside}. Many papers have contributed to profiling partisan news spreaders and decreasing their impact on society. 
Karamshuk \etal{}~\cite{karamshuk2016identifying} study the linguistic choices for political agendas, and propose a natural language processing algorithm to identify partisan bias for Twitter users.
Recently, Sakketou \etal{}~\cite{sakketou2022factoid} introduced the first Reddit dataset targeting users who spread fake partisan news, namely FACTOID. This dataset captures users' historical posts and interaction data and is validated by a psycho-linguistic feature analysis for bias classification. In this work, we dive into the role of spreaders' partisan bias in influencing their communicative activities. Building on previous work~\cite{garimella2018political}, our analysis investigates features in communication and reveals that, during the invasion, spreaders with highly partisan bias communicate differently from those with a more neutral bias.

\section{Methodology}
\label{sec:methodology}


In this section, we first explain the data collection, followed by the terminology used in this paper, data annotation and partisan bias calculation. 

\subsection{Dataset}

\begin{table*}[t]
\centering
\resizebox{.95\textwidth}{!}{%
\begin{tabular}{lrrrrrr}
\toprule
\textbf{Subreddit} & \textbf{\#Submissions} & \textbf{\#Users} & \textbf{\#URLs} & \textbf{\#Submission with URLs} & \textbf{\#PSubmission} & \textbf{\#PCommunicator} \\
\midrule
\textit{russia}                & 86,423    & 16,758    & 9,300    & 5,432   & 506    & 272    \\[0.4ex]
\textit{ukraine}               & 4,094,867 & 262,437 & 293,246 & 181,850 & 27,136 & 11,206 \\[0.4ex]
\textit{ukraina}               & 146,560   & 25,978    & 13,510    & 7,015   & 615    & 389    \\[0.4ex]
\textit{RussiaUkraineWar2022}  & 416,005 & 48,850   & 17,139   & 9,904 & 1,732  & 1,121  \\[0.4ex]
\textit{UkraineWarVideoReport} & 1,529,658 & 156,889   & 89,048   & 53,460  & 7,877  & 4,223  \\[0.4ex]
\textit{UkrainianConflict}     & 1,802,330   & 129,155   & 106,606    & 71,610  & 15,053 & 5,867  \\ 
\bottomrule
\end{tabular}%
}
\caption{Description of each subreddit in the collected dataset. The \enquote{\#} means \enquote{the count of}, \ie{\#Submissions denotes the count of submissions.}
}
\label{tab:dataset}
\end{table*}

\update{We select six representative subreddits related to the Russian invasion, as suggested by \cite{RUDataset}.
Three subreddits --- \textit{russia}, \textit{ukraine}, \textit{ukraina} --- are directly related to Russia and Ukraine, the two warring sides of the Russian Invasion 2022. As reported in~\cite{tkachenko2019conflict, pierri2022does}, these three subreddits produce event-oriented conversation during the ongoing political conflict between Ukraine and Russia. In addition, recent literature has uncovered the widespread sharing of partisan news content in these subreddits~\cite{hanley2022happenstance}. Thus, we consider these three subreddits suitable sources to conduct our analysis.}

\update{For other three subreddits --- \textit{RussiaUkraineWar2022} (193K members), \textit{UkraineWarVideoReport} (640K members), and \textit{UkrainianConflict} (426K members) --- we select them because they are the largest
Reddit news-sharing communities oriented towards the war. News content and media shared in these three subreddits are therefore more focused on the ongoing invasion. 
In addition, related studies have shown that partisan news is more likely to be spread in such communities oriented by ongoing political events~\cite{10.1145/3131365.3131390, 10.1145/3342220.3343662}. Thus, we include these three subreddits.}


We use the Pushshift API \cite{baumgartner2020pushshift} to extract all submissions (posts and comments) from January 1\textsuperscript{st} to August 31\textsuperscript{st} of 2022. Note, we filter out submissions without any text or containing \enquote{[deleted]} or \enquote{[removed]} labels.
We also remove potential bot accounts whose usernames contain \enquote{bot}, \enquote{Bot}, \enquote{auto} or \enquote{Auto}~\cite{li2022all}. Our dataset contains 8,075,843 submissions. During the eight months, 468,824 users have posted text and shared 528,849 URLs in total. Table~\ref{tab:dataset} summarizes the statistics for all subreddits in the dataset.



After removing bot accounts, we identify 52,919 (0.65\%) submissions sharing URLs from partisan news (from an overall set of 329,271 submissions containing URLs).
We identify these using a verified list (see Section~\ref{sec:methodology:partisan degree}). 
These submissions have accumulated 211,796 comments and contain a total of 77,871 (14.72\%) URLs from 933 distinct partisan news domains.

\subsection{Computing partisan score}
\label{sec:methodology:partisan degree}

\begin{figure}[]
  \centering
  \includegraphics[width=\columnwidth]{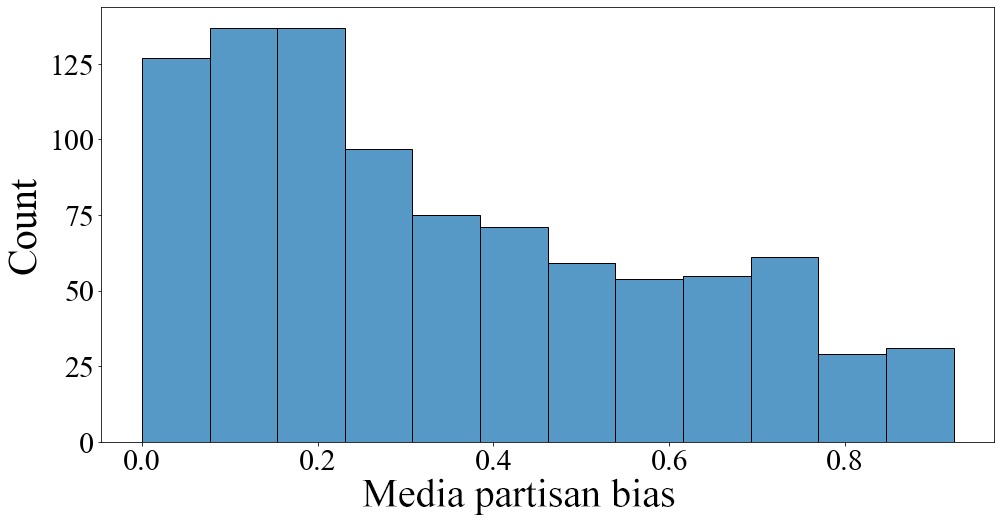}
  \caption{Histogram of media partisan bias for news media in our dataset.}
  \label{fig:partisan_score}
\end{figure}

We next tag URLs with a partisan score, as proposed in~\cite{garimella2018political}.



\pb{News annotations.}
We first extract the news outlets from all URLs in the dataset. To do so, we use a media list from Media Bias/Fact Check (MBFC) (as of October 2022). MBFC is a fact-checking platform offering political bias ratings for news sources.\footnote{\url{https://mediabiasfactcheck.com/methodology/}} 
This list contains 1,703 verified news domains, out of which 933 domains are present in our dataset.  We only use these 933 news domains for the rest of the analysis in this paper. 

\pb{Partisan score.}
For measuring the media source's partisan bias, we use the absolute value of Robertson's partisan score~\cite{robertson2018auditing}. The bias ranges from 0 (not partisan) to 1 (very partisan). Figure~\ref{fig:partisan_score} shows the distribution of news media's partisan bias in our dataset. We assign the corresponding bias from our list to each URL in our dataset.

Based on this, we then assign a \textit{partisan score} to each user, quantifying how biased they are when spreading partisan news. A higher partisan score means that the corresponding user focuses on spreading news media with a highly partisan bias. For a user, their personal partisan score is simply the average of all annotated URLs' partisan scores they have shared.
Specifically, for a given user $x$, $L_x$ is the set of annotated URLs shared by $x$; and the partisan score $d(x)$ is the average of all annotated URLs’ partisan bias:
\[d(x) = \frac{1}{|L_{x}|} \sum partisan~bias(l_x),\quad l_x \in L_x\]

\subsection{Definitions}

We use the word \enquote{submission} for both posts and comments on Reddit. 
\begin{itemize}
    \item \textbf{PSubmission} refers to the submissions containing news URLs annotated with partisan score.  
    
    \item \textbf{PCommunicators} refers to users who submit the submissions with URLs annotated with partisan score, \ie~PSubmissions. 
    
\end{itemize}

\subsection{Ranking PCommunicators}
As mentioned in~\cite{garimella2018political}, a PCommunicator can be
\one~\textit{partisan} (focused on spreading media with highly partisan bias); 
or
\two~\textit{neutral} (focused on spreading media with neutral bias). 
We want to investigate whether PCommunicators present differences in communication if they produce news content from media with a high partisan bias.
Referring to previous work~\cite{garimella2018political}, we define a PCommunicator as $N^{th}$-partisan, for integer $0<N\leq50$, if their partisan score is above $N$ percentile on the distribution of partisan score from total PCommunicators. 
For the remaining PCommunicators who are not $N^{th}$-partisan, we call them $N^{th}$-neutral. \update{To provide an example, say that $N=30$. In this case, PCommunicators would be considered partisan (or $30^{th}$-partisan) only when their partisan score exceeds the $30^{th}$ percentile of partisan scores for the entire group of PCommunicators. Conversely, PCommunicators would be deemed neutral (or $30^{th}$-neutral) if their partisan score falls below the $30^{th}$ percentile.}
\section{Partisan News Sharing in Subreddits}

\update{In this section, we investigate how the sharing of partisan news evolves during the war. We examine the temporal distribution of PSubmissions and their received comments to understand how users' communication activities change during the war period. In addition, we inspect the relation between a news article's partisan bias and the number of subreddits it spreads across.}

\subsection{Temporal Evolution of Partisan News}

\label{sec:Temporal Evolution}
\begin{figure}[]
  \centering
  \includegraphics[width=\columnwidth]{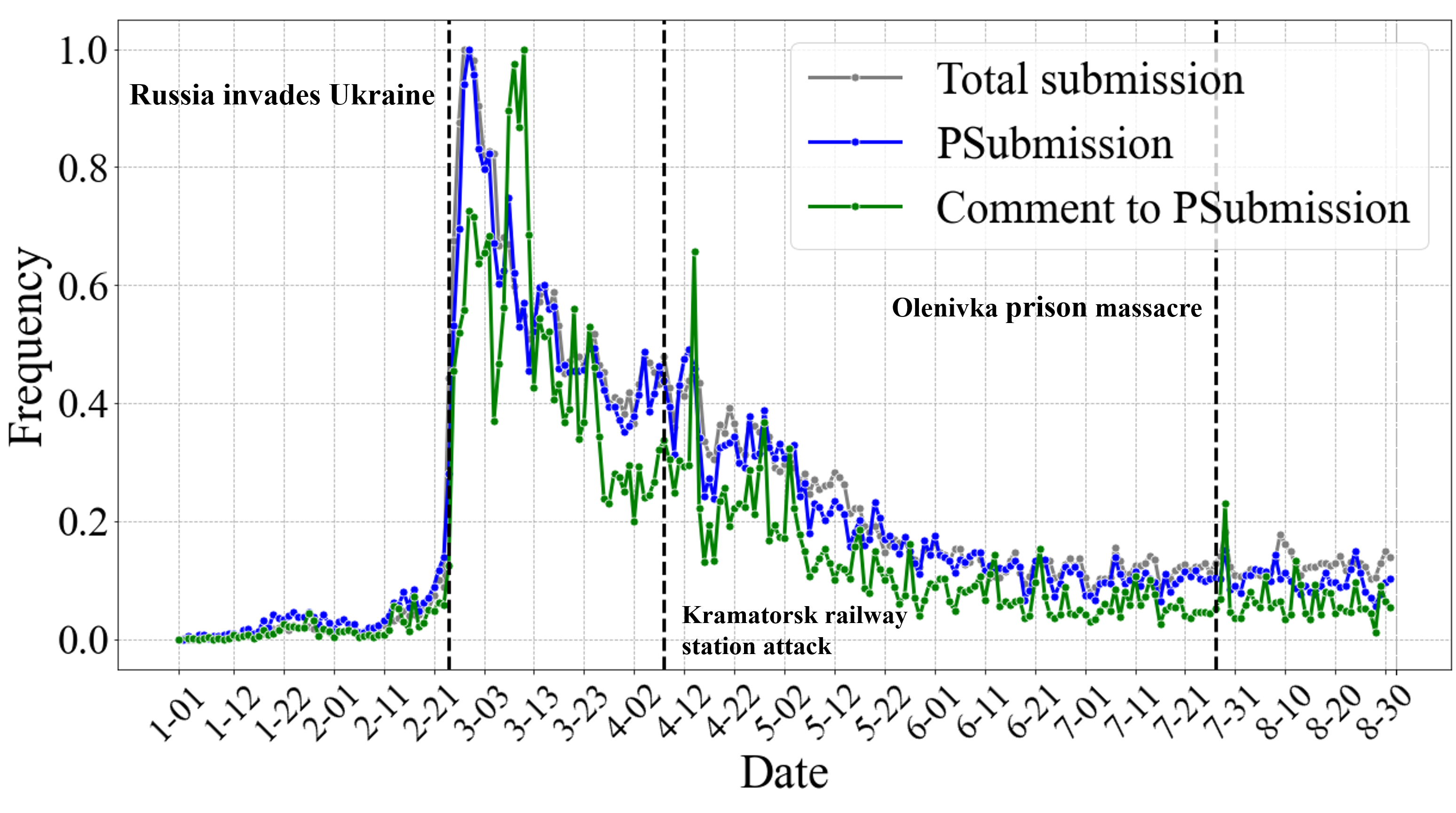}
  \caption{Max-min normalization frequency of total submissions, PSubmissions and PComments.}
  \label{fig:temporal_distribution}
\end{figure}

We first hypothesize that there would be an increase in the sharing of partisan news URLs during major events of the Russian invasion. 
We further conjecture that such an increase will attract more users to get involved in the discussion by submitting more comments.
Figure \ref{fig:temporal_distribution} depicts a time series of the number of \textit{total submissions}, \textit{PSubmissions}, and \textit{the comments to PSubmissions}.
Note, the time series for these items is based on the date when they are submitted to Reddit.

There is a surge in the daily volume of submissions after the launch of the Russian invasion. The max-min normalized frequency increases from 0.15 (February \(23^{rd}\)) to 1.0 (February \(26^{th}\)), indicating a large increase in activities among the selected subreddits. As expected, we also find that both PSubmissions and the comments to PSubmissions follow a similar pattern to the total submissions and have a higher daily frequency compared to the pre-invasion period (before February \(24^{th}\)). 

We observe three significant peaks in the comments' evolution following the key events during the invasion. The first peak is after the Russian occupation of Kherson Oblast (March \(2^{nd}\)),
the second one is after the Kramatorsk railway station attack (April \(8^{th}\)) and the third one is after the Olenivka prison massacre (July \(29^{th}\)).
Such variance reflect the sensitivity of comments to breaking events:
commenting behaviors to the partisan news increase following the breaking events. 

\subsection{Partisan News Sharing across Multiple Subreddits} 
\label{sec:range}



\begin{figure*}[]
    \centering
  \begin{subfigure}[]{.98\columnwidth}
    \centering
    \includegraphics[width=.98\columnwidth]{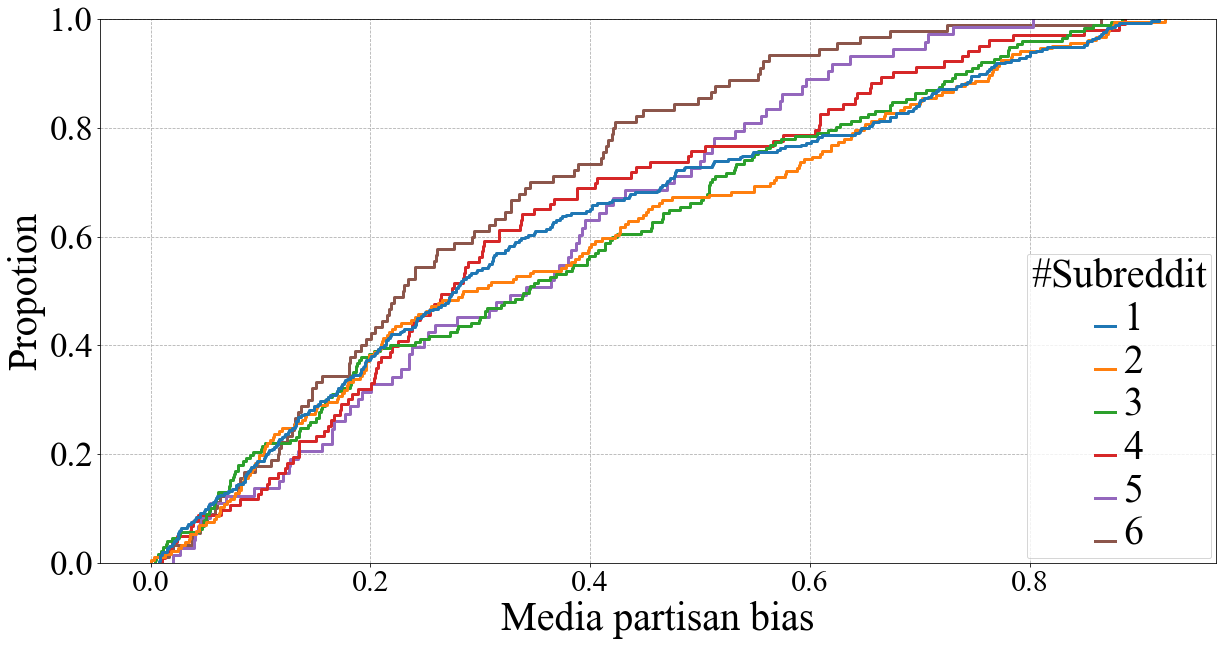}
    \caption{All domains in each groups.}
    \label{fig:cross_sub_all}
  \end{subfigure}
  \hfil
  \begin{subfigure}[]{.95\columnwidth}
    \centering
    \includegraphics[width=.95\columnwidth]{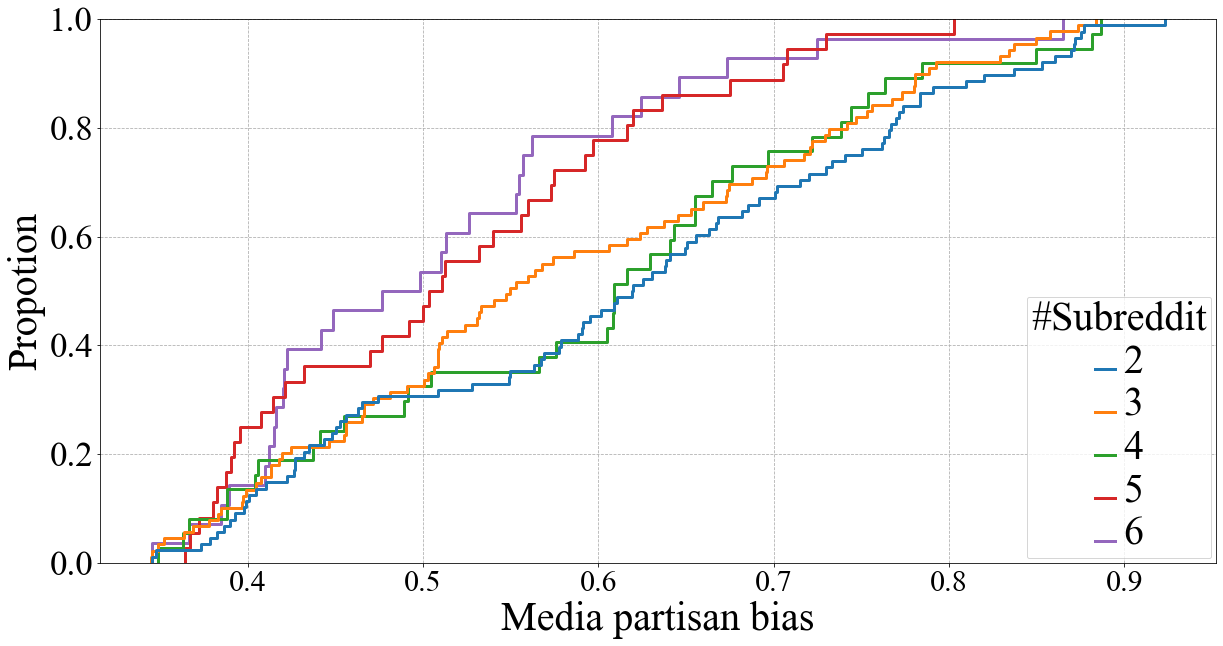}
    \caption{Multi-crossing domains with above 50\% partisan score in each groups.}
    \label{fig:cross_sub_50}
  \end{subfigure}
  \caption{The distribution of partisan score. The ``\#Subreddit'' denotes the number of subreddit that the group of partisan news have appeared in. A cdf-line posits above means the group of partisan news hold a lower distribution of paritsan bias.
  }
  \label{fig:cross_sub}
\end{figure*}

\begin{table}[]
\centering
\resizebox{0.85\columnwidth}{!}{%
\begin{tabular}{ccccccc}
\toprule
                      & \multicolumn{6}{c}{\textit{\textbf{Y}}}                        \\[0.4ex] \midrule
\multicolumn{1}{c|}{\multirow{6}{*}{\textit{\textbf{X}}}} & \multicolumn{1}{c|}{\textbf{\#Subreddit}} & 2 & 3 & 4 & 5 & 6 \\[0.4ex] \cline{2-7}\\[-2ex]
\multicolumn{1}{c|}{} & \multicolumn{1}{c|}{2} & -     & 0.046  & 0.502 & 0.002   & ***   \\[0.4ex]
\multicolumn{1}{c|}{} & \multicolumn{1}{c|}{3} & 0.845 & -     & 0.845 & 0.039  & 0.048   \\[0.4ex]
\multicolumn{1}{c|}{} & \multicolumn{1}{c|}{4} & 0.783 & 0.136  & -     & 0.004 & 0.002 \\[0.4ex]
\multicolumn{1}{c|}{} & \multicolumn{1}{c|}{5} & 0.960 & 0.814 & 0.864 & -     & 0.583 \\[0.4ex]
\multicolumn{1}{c|}{} & \multicolumn{1}{c|}{6} & 1.0   & 0.942   & 0.890   & 0.510 & -     \\ \bottomrule
\end{tabular}%
}
\caption{P-values of two-sample Kolmogorov–Smirnov
test (one-sided) on domain partisan score for each group pair. The alternative hypothesis is that \(X>Y\) and the number in the cell refer to the p-value for the comparison between the two groups. *** denotes that \(p<0.001\).} 
\label{tab:2sampleks}
\end{table}

We next investigate the relationship between the partisan bias and the number of subreddits the news domain is shared across. 
We consider that the more subreddits a domain has appeared, the more widely it is propagated during the Russian invasion. 

We first group partisan news domains by how many subreddits they have appeared in (\textit{\#Subreddit}).
Figure \ref{fig:cross_sub_all} plots the cumulative distribution
of domains' partisan scores for each group. We observe that, when it comes to low partisan bias, there is no obvious stratification in the distribution between each group.
However, as the partisan bias increases
the stratification becomes clearer. Specifically, the CDF-lines with higher \#Subreddit are higher in the plot. This indicates that groups crossing more subreddits (greater \#Subreddit) hold lower partisan bias.
In other words, highly partisan news is \emph{less} likely to be spread across multiple subreddits, indicating the role of media partisan bias in limiting the number of subreddits that news domains spread across.

To verify our conclusion, we extract domains that have been posted on multiple subreddits (\(\#Subreddit > 1\)) with a partisan bias above the average (\(\mu=0.344\)).
Figure \ref{fig:cross_sub_50} plots the distribution of partisan bias for these partisan domains. Aligned with our expectations, we find that more partisan domains are posted across fewer subreddits.
We then apply a one-side two-sample Kolmogorov–Smirnov test on each group pair to see whether domains crossing more subreddits hold lower partisan bias distribution.
Table \ref{tab:2sampleks} shows the p-values of test results. Except the \#Subreddit pair (2 \vs{4}), (3 \vs{4}) and (5 \vs{6}), all other groups pairs (\textit{X} \vs{\textit{Y}}) accept the alternative hypothesis (\(p<0.05\)) when \#Subreddit(\textit{X}) \(<\) \#Subreddit(\textit{Y}).
In this case, X holds significantly higher partisan bias than Y. These results confirm our above observation that, as the partisan bias increases, it is less likely for a new domain to be spread in multiple subreddits. 
\update{Our analysis suggests that media with a neutral stance tends to be spread more broadly than those with a high bias. }




\section{Communities Response to PSubmissions}
\label{sec:response}

We now explore how PSubmissions  are responded to by other users. As shown in~\cite{karami2021profiling}, a user may be encouraged to submit more partisan news if they receive a more active response from other users. Therefore, we check if PSubmissions' receive more comments and upvotes.


\subsection{Metrics}

We focus on the following two metrics to gauge the response to a submission.

\pb{Normalized number of comments:} As shown in \cite{aldous2019view, risch2020top}, the number of comments for a submission is an effective indicator for engagement. We utilize the normalized number of comments to measure users' \textit{engagement} in the discussion. The number of comments refers to the number of comments on a Reddit post or comment. The normalized number of comments is a normalization of comments' number for a submission by the number of users in the corresponding subreddit. A higher normalized number of comments indicates more users' engage with a submission.


\pb{Normalized voting score:} 
Users can vote submissions up or down to show their reception of content. The public voting trend can affect the topic's leaning in a subreddit \cite{glenski2017consumers}. The \textit{voting score} refers to the number of upvotes minus the number of downvotes for a submission. 
As reported by Waller \etal{} \cite{waller2019generalists}, these voting scores are a proxy of public \textit{reception} to media-related content.
We utilize normalized voting scores to measure communities' reception of media links. The normalized voting score is a normalization of a submission's score by the number of users in the corresponding subreddit.
A higher normalized voting score indicates higher reception of a submission's content. 

Note, the normalization is performed to remove the influence introduced by the size of the subreddits. Submissions on larger subreddits are more likely to be viewed by more users and therefore receive more comments and votes. 

\subsection{Comparative Analysis}
\label{sec:Comparison Analysis}


Using the above metrics, we test whether PSubmissions receive a more active response -- more comments and upvotes --- than other types of submissions, \ie{ submissions without URLs
or sharing URLs but not from partisan news}. 

For comparison, we classify all submissions based on the URLs contained in the text using three categories:
\one~\textbf{Non-URL}: submissions without any URLs;
\two~\textbf{Non-annotated}: submissions with URLs but not annotated as being partisan news;
and 
\three~\textbf{PSubmission}: submissions with URLs from partisan news. 
Using these groups, we inspect whether there is 
a significant difference in 
normalized number of comments and voting scores.
Note, we divide the posts and comments into two groups and perform the comparison analysis separately.



We first examine the distribution of the two metrics for each group. A normality test reports that our samples do not follow a normal distribution. For posts, a Shapiro-Wilk Test reveals a shred of significant evidence that normalized number of comments (\(\mu=3.42e^{-4},\delta=1.29e^{-3},Md=8.8e^{e-6},min=0,max=0.056\)) does not come from a normal distribution (\(W=0.198,p<0.001\)).
Thus, to inspect the inter-group difference, we perform a Kruskal Wallis test with a Dunn’s post-hoc test on normalized number of 
comments and voting score independently. Table~\ref{tab:KW test} shows our statistics results, and we detail our findings as follows. 



\update{For received comments, the post-hoc test shows that both \textbf{posts} and \textbf{comments} in PSubmissions 
hold a significantly (\(p<0.001\)) higher normalized number of comments than in Non-URL 
and Non-annotated.
This indicates that users are more likely to comment on PSubmissions when compared to submissions without media sharing if they contain non-partisan media. Thus, sharing partisan news does attract more users to engage in discussion. This is intuitive as inclusion of news URLs increases the entropy of a submission.}

\update{For submissions' voting scores,
the post-hoc test shows that only \textbf{posts} in PSubmissions
receive a significantly (\(p<0.001\)) higher normalized voting score than in both Non-URL
and Non-annotated.
As for the \textbf{comments}, we find PSubmissions
receive a significantly (\(p<0.001\)) lower normalized voting score compared to both Non-URL and Non-annotated.
The test results suggest that users are more willing to upvote a post with partisan news, indicating a higher public reception to posts' content.
Commenting with partisan news, however, cannot increase reception to comments' content.} 


In summary, sharing partisan news attracts users to engage in the discussion, submitting more comments to the PSubmissions. Users are also more likely to upvote posts with URLs from partisan news. This indicates that sharing partisan news can increase communities' reception of a post's content. However, comments sharing partisan news seem to receive much lower voting scores.

\begin{table*}[t]
    \centering
        \centering
        \resizebox{.9\textwidth}{!}{
        \begin{tabular}{lc|lc}
            \toprule
             \textbf{Metrics} & \textbf{KW H} & \textbf{Mean diff (post-hoc)}       & \textbf{\textit{p} (post-hoc)} \\[0.4ex] \midrule
            \multirow{3}{*}{Normalized number of comments (post)} & \multirow{3}{*}{\begin{tabular}[c]{@{}c@{}}305.85\\ (***)\end{tabular}}  & Non-annotated ($2.96e^{-4}$) \textless Non-URL ($3.69e^{-4}$)    & *** \\[0.4ex]
             &               & PSubmission ($3.95e^{-4}$) \textgreater Non-URL ($3.69e^{-4}$)       & ***                 \\[0.4ex]
             &               & PSubmission ($3.95e^{-4}$) \textgreater Non-annotated ($2.96e^{-4}$) & ***                 \\[0.4ex] \midrule
            \multirow{3}{*}{Normalized voting score (post)} & \multirow{3}{*}{\begin{tabular}[c]{@{}c@{}}1387.91\\ (***)\end{tabular}} & Non-annotated ($1.50e^{-3}$) \textgreater Non-URL ($1.00e^{e-3}$) & *** \\[0.4ex]
             &               & PSubmission ($1.90e^{-3}$) \textgreater Non-URL ($1.00e^{e-3}$)       & ***                 \\[0.4ex]
             &               & PSubmission ($1.90e^{-3}$) \textgreater Non-annotated ($1.50e^{-3}$) & ***                 \\[0.4ex] \midrule
             \multirow{3}{*}{Normalized number of comments (comment)} & \multirow{3}{*}{\begin{tabular}[c]{@{}c@{}}2925.50\\ (***)\end{tabular}} & Non-annotated ($8.50e^{-6}$) \textless Non-URL ($8.59e^{-6}$)    & *** \\[0.4ex]
         &               & PSubmission ($1.12e^{-5}$) \textgreater Non-URL ($8.59e^{-6}$)       & ***                 \\[0.4ex]
         &               & PSubmission ($1.12e^{-5}$) \textgreater Non-annotated ($8.50e^{-6}$) & ***                 \\[0.4ex] \midrule
        \multirow{3}{*}{Normalized voting score (comment)} & \multirow{3}{*}{\begin{tabular}[c]{@{}c@{}}3208.11\\ (***)\end{tabular}} & Non-annotated ($5.07e^{-5}$) \textless Non-URL ($6.36e^{-5}$) & *** \\[0.4ex]
         &               & PSubmission ($4.84e^{-5}$) \textless Non-URL ($6.36e^{-5}$)      & ***                 \\[0.4ex]
         &               & PSubmission ($4.84e^{-5}$) \textless Non-annotated ($5.07e^{-5}$) & ***                 \\[0.4ex] 
         \bottomrule
        \end{tabular}}
    
    
    \caption{Pair-wise comparison of posts/comments groups in PSubmissions by the Kruskal Wallis test with Dunn’s post-hoc test on normalized number of comments and normalized voting score. \enquote{Mean diff} column shows the comparison results of the mean value of corresponding metrics between two groups. *** denotes that \(p<0.001\).}
    \label{tab:KW test}
\end{table*}

\section{Profiling Partisan Communicators}
\label{sec:PCommunicator}


In this section, we analyse the activities of PCommunicators. 
Recall that we rank PCommunicators as $N^{th}$-partisan who focus on producing news content with highly partisan bias.
$N^{th}$-neutral refers to PCommunicators spreading news media with a more neutral bias. 
In this part, we aim to explore how partisan PCommunicators act differently from bipartisan PCommunicators in communication.

\subsection{Features in Communication}
\label{sec:Metrics of Activities}

\update{We limit our analysis to non-deleted users in our dataset. This decision is made because the Pushshift API replaces the names of all deleted users with the label ``[deleted]'', which makes it impossible to distinguish between the identities of different deleted users based on their names.}
We also focus on users who have contributed 3 ore more PSubmissions (number of PSubmissions \(\geq3\)). 
We study these users' activities from two perspectives, detailed below.

\pb{Content production:}
We inspect how much partisan content is produced by these users. We utilize two features to characterize users' sharing of partisan news items:
\begin{itemize}
    \item The number of PSubmissions by a user.
    \item The average number of annotated URLs per PSubmission.
\end{itemize}


\pb{Community response:} 
We also inspect the community response
received by users who spread partisan news. 
We use four features to characterize the response of PCommunicators:
\begin{itemize}
    \item The average normalized number of comments per PSubmissions.
    \item The average normalized voting score per PSubmissions.
    \item The average normalized number of distinctive commenters per PSubmissions.
    \item The proportion of commented PSubmissions (\ie comment rate).
\end{itemize}

\subsection{Analysis of Features in Communication}
\label{sec:metrics result}

\begin{table*}[t]
\centering
\resizebox{.8\textwidth}{!}{%
\begin{tabular}{lc|rrc}
\toprule
\textbf{Metrics} & \textbf{Significance} & \textbf{coef} & \multicolumn{1}{l}{\textbf{std err}} & \textit{\textbf{p}} \\ \midrule
(Intercept) &           & $8.28e^{-1}$     & $1.56e^{-1}$   & ***   \\[0.4ex]
Number of PSubmissions &\cmark      & $-1.60e^{-3}$     & $1.00e^{-3}$   & 0.190 \\[0.4ex]
Average number of annotated URLs &\cmark       & $4.44e^{-1}$     & $9.10e^{-1}$   & ***   \\[0.4ex]
Average of normalized number of comments (post) &       & $-2.02e^{+5}$    & $2.48e^{+4}$ & ***   \\[0.4ex]
Average of normalized number of commenters (post) &   & $4.97e^{+5}$    & $6.10e^{+4}$  & ***   \\[0.4ex]
Average of normalized voting score (post) &\cmark (-)       & $-6.09e^{+3}$    & $7.86e^{+2}$   & ***   \\[0.4ex]
Comment rate (post)  &          & $1.94e^{+0}$   & $3.83e^{-1}$ & ***   \\[0.4ex]
Average of normalized number of comments (comment) &\cmark     & $4.85e^{+5}$    & $2.69e^{+4}$ & ***   \\[0.4ex]
Average of normalized number of commenters (comment) &\cmark & $1.34e^{+5}$   & $4.65e^{+4}$ & **   \\[0.4ex]
Average of normalized voting score (comment)  &   & $-1.81e^{+4}$    & $1.21e^{+3}$   & ***   \\[0.4ex]
Comment rate (comment) &\cmark         & $-6.96e^{+0}$    & $3.67e^{-1}$ & ***  \\[0.4ex] \midrule
Log-Likelihood   & \multicolumn{4}{c}{-7246.2}  \\[0.4ex]
AIC              & \multicolumn{4}{c}{14514.48} \\[0.4ex]
MAE              & \multicolumn{4}{c}{0.3585} \\[0.4ex]
Accuracy         & \multicolumn{4}{c}{64.15\%}  \\[0.4ex] 
\bottomrule
\end{tabular}%
}
\caption{\update{A summary of significance on metrics to profile partisan PCommunicators and performance of Logistic Regression model to predict partisan PCommunicators. We experimented with six values of the threshold \(N\). In the ``Significance'' column, a \enquote{\cmark} (\enquote{\cmark (-)}) denotes that the corresponding metric is significantly higher (lower) in $N^{th}$-partisan than $N^{th}$-neutral PCommunicators (\(p<0.05\)), for \textit{at least four out of six} thresholds \(N\) experimented.}}
\label{tab:Logistic Regression}
\end{table*}

In this part, we study which of the aforementioned metrics significantly differentiate partisan and neutral PCommunicators.
Referring to~\cite{garimella2018political}, our ranking for partisan and neutral is parameterized by the percentile threshold \(N\) on PCommunicators' partisan score. We experiment with different values of \(N\) as in [25, 30, 35, 40, 45, 50].
For each value of \(N\), we examine the distribution of the above features in the $N^{th}$-partisan and $N^{th}$-neutral, and test if there are any significant inter-group differences. For comparison, we apply Mann–Whitney U test on each features respectively. 

Table~\ref{tab:Logistic Regression} summarizes the features that are significantly different between the two groups, in a majority of the experiments on diverse \(N\) values. \update{A \enquote{\cmark} (\enquote{\cmark (-)}) means that the corresponding features (\ie{number of PSubmissions}) is significantly (\(p<0.05\)) higher (or lower) in $N^{th}$-partisan than $N^{th}$-neutral for \textit{at least four out of six} $N$ thresholds.}
Our test finds \update{six} features with a significant difference in their distributions between partisan and neutral PCommunicators.

For the content production metrics, partisan PCommunicators submit significantly more PSubmissions than neutral PCommunicators. In addition, partisan PCommunicators also tend to contain more URLs from partisan news when writing PSubmissions. These results suggest that, compared to neutral PCommunicators, partisan PCommunicators are more active in generating discussion on partisan news and produce much more partisan-related content individually.


\update{For the community response metrics, our findings reveal different outcomes when examining posts vs.\ comments. 
In regards to \textit{posting}, partisan PCommunicators tend to receive significantly lower voting scores compared to neutral PCommunicators. This suggests that in the observed communities, sharing news media with highly partisan leaning in posts is less receivable than sharing neutral ones. On the other hand, when \textit{commenting} with partisan news, partisan PCommunicators receive a significantly higher volume of comments from more distinctive commenters and hold a higher comment rate than neutral PCommunicators. This suggests that by sharing media with highly partisan bias in comments, partisan PCommunicators are more likely to attract more users to engage in the discussion.}


\subsection{Prediction}

The above leads us to perform logistic regression to better understand the key features that distinguish partisan and neutral users.
That is, we formulate a predictive task using the above features. We formulate a Logistic Regression model for this task as:
\[P(Y_u=1)=\beta_0+\sum_{x_i \in X}{\beta_i x_i}+\epsilon\]
where \(Y_u\) denotes the possibility of a PCommunicator \(u\) to be partisan (\(Y_u=1\)) or neutral (\(Y_u=0\)). The \(X\) refers to the set of metrics we define in Section~\ref{sec:Metrics of Activities}.


To test our classifier, we choose an intermediate experimented threshold \(N=30\) to label the PCommunicators as partisan or neutral. 
We randomly split our data into training and testing sets by a ratio of 80:20. 
\update{We repeat the splitting and testing 100 times to validate our classifier's performance. 
In all, our classifier achieves an average accuracy 61.57\% ($SD=1.14\%$) and holds an average mean absolute error (MAE) of 0.3844 ($SD=0.0114$). 
Table~\ref{tab:Logistic Regression} reports the regression coefficients and the performance of the model with the highest accuracy (64.15\%).
The \enquote{\cmark} (\enquote{\cmark (-)}) means that the corresponding features are significantly (\(p<0.05\)) higher (lower) for partisan PCommunicators, as examined in Section~\ref{sec:metrics result}.} 
\update{The results confirm that, except the number of individual submitted PSubmissions, all features have statistical significance in correlating with a PCommunicator being partisan.
We argue this can also be used to facilitate moderators in identifying such patterns.}


\section{Analyzing the Comment Network of PCommunicators}
\label{sec:network}
In this section, we explore the network properties of PCommunicators by analyzing the interaction networks driven by their commenting behavior. We again use the $N^{th}$-partisan division to distinguish between partisan and neutral PCommunicators. We aim to determine if there are any significant differences in network positions for these two groups.

\subsection{Generating the Comment Network}

We induce a comment network~\cite{nadiri2022large} to represent the connections among users in Reddit. In this network, nodes represent users, and edges indicate comment interactions between them. To focus on the commenting behaviors associated with the sharing of partisan news, we only consider comments generated by PSubmissions. If a user leaves a comment under a PSubmission, we assign a \textit{directed} link from this user to the PCommunicator (author of that PSubmission). Each link is weighted by the total number of comment between the user and the PCommunicator. The resulting comment network is a directed graph consisting of 50,162 nodes and 94,105 edges, where these connections are generated by 20,294 PCommunicators. 

\subsection{Centrality Features of PCommunicators}


Existing studies have proposed that sharing partisan news can impact the spreaders' influence on their social networks~\cite{garimella2018political, flamino2023political}. Therefore, we examine the network features of PCommunicators to investigate if they demonstrate different levels of network influence. To explore this, we investigate various notions of centrality to examine PCommunicators' network influence~\cite{10.1093/acprof:oso/9780199206650.001.0001}:





\pb{In-degree centrality:}
The in-degree centrality of a node in a comment network is calculated by counting the number of incoming links that connect to that node. As noted in~\cite{bian2019identifying}, in-degree centrality can be interpreted as a measure of popularity. In the context of our study, a higher in-degree centrality for PCommunicators indicates that they have gained more popularity by sharing partisan news, as they are more likely to receive comments from other users in the network.


\pb{Betweenness centrality:}
The betweenness centrality is a measure of the number of times a node serves as a bridge along the shortest path between two other nodes. It can also identify brokerage positions in networks that connect different communities~\cite{freeman2002centrality}. As for a comment network, a higher betweenness centrality suggests that PCommunicators act as connectors between communities by sharing partisan news, and they help to form connections between users who would not otherwise interact.

\pb{PageRank:}PageRank~\cite{page1998pagerank} is a variant of the eigenvector centrality that assigns a score to each node based on the importance of its neighbors. In a comment network, a PCommunicator with a higher PageRank can be interpreted as being more influential in the network. As suggested by~\cite{herzig2014author}, this means that their shared partisan news are more significant or more likely to be commented by other influencers in the network.

\subsection{Analysis of Centrality Features}


\begin{table*}[]
\centering
\resizebox{\textwidth}{!}{%
\begin{tabular}{lcccccc}
\toprule
\multirow{2}{*}{\textbf{Metrics}} &
  \multicolumn{6}{c}{\textbf{Mean diff (Mann–Whitney U) for PCommunicaors}} \\[0.4ex] \cline{2-7}\\[-6px]
 &
  \textbf{N=25} &
  \textbf{N=30} &
  \textbf{N=35} &
  \textbf{N=40} &
  \textbf{N=45} &
  \textbf{N=50} \\
  \midrule
In-degree centrality &
  \begin{tabular}[c]{@{}c@{}}neutral \textgreater partisan\\ (***)\end{tabular} &
  \begin{tabular}[c]{@{}c@{}}neutral \textgreater partisan\\ (***)\end{tabular} &
  \begin{tabular}[c]{@{}c@{}}neutral \textgreater partisan\\ (***)\end{tabular} &
  \begin{tabular}[c]{@{}c@{}}neutral \textgreater partisan\\ (0.079)\end{tabular} &
  \begin{tabular}[c]{@{}c@{}}neutral \textgreater partisan\\ (0.405)\end{tabular} &
  \begin{tabular}[c]{@{}c@{}}neutral \textgreater partisan\\ (0.970)\end{tabular} \\[8px]
Betweenness centrality &
  \begin{tabular}[c]{@{}c@{}}neutral \textless partisan\\ (***)\end{tabular} &
  \begin{tabular}[c]{@{}c@{}}neutral \textless partisan\\ (***)\end{tabular} &
  \begin{tabular}[c]{@{}c@{}}neutral \textless partisan\\ (***)\end{tabular} &
  \begin{tabular}[c]{@{}c@{}}neutral \textless partisan\\ (0.324)\end{tabular} &
  \begin{tabular}[c]{@{}c@{}}neutral \textgreater partisan\\ (0.816)\end{tabular} &
  \begin{tabular}[c]{@{}c@{}}neutral \textgreater partisan\\ (0.398)\end{tabular} \\[8px]
PageRank (\cmark (-)) &
  \begin{tabular}[c]{@{}c@{}}neutral \textgreater partisan\\ (***)\end{tabular} &
  \begin{tabular}[c]{@{}c@{}}neutral \textgreater partisan\\ (***)\end{tabular} &
  \begin{tabular}[c]{@{}c@{}}neutral \textgreater partisan\\ (***)\end{tabular} &
  \begin{tabular}[c]{@{}c@{}}neutral \textgreater partisan\\ (0.043)\end{tabular} &
  \begin{tabular}[c]{@{}c@{}}neutral \textgreater partisan\\ (0.419)\end{tabular} &
  \begin{tabular}[c]{@{}c@{}}neutral \textgreater partisan\\ (0.994)\end{tabular} \\[8px] \bottomrule
\end{tabular}%
}
\caption{A summary of significance in test results on various network metrics between partisan and neutral PCommunicators. We experimented with six values of the threshold \(N\). \enquote{\cmark (-)} denotes that the corresponding metric is significantly lower in $N^{th}$-partisan than $N^{th}$-neutral PCommunicators (\(p<0.05\)), for \textit{at least four out of six} thresholds \(N\) experimented. \enquote{Mean diff} column shows the comparison results of the mean value of corresponding metrics between neutral and partisan PCommunicators. *** denotes that \(p<0.001\).}
\label{tab:net_metrics}
\end{table*}

We focus our analysis on non-deleted users who have contributed multiple PSubmissions (\#PSubmissions \(\geq3\)). We experiment with different values of $N$ as in [25, 30, 35, 40, 45, 50] and examine the distribution of the network features in $N^{th}$-partisan and $N^{th}$-neutral PCommunicators. We apply the Mann–Whitney U test to test if there are any significant inter-group differences. Table~\ref{tab:net_metrics} summarizes the network features that significantly differ between the two groups \textit{for at least four out of six} experimented values of $N$. 

We only observe that PageRank has a significant difference in its distributions between partisan and neutral PCommunicators. Partisan PCommunicators tend to have significantly lower PageRank compared to neutral PCommunicators. 
Related studies have shown that a user's social influence on Reddit can be estimated using the PageRank on interaction networks, with the received voting score serving as a proxy~\cite{stoddard2015popularity, massachs2020roots}. Based on this assumption, we conduct a Pearson correlation test between the PageRank of PCommunicators and their average normalized voting score 
per post (as examined in Section~\ref{sec:metrics result}). 
The result shows a significant positive correlation between these two metrics ($r=0.3718, p<0.001$), suggesting that a higher PageRank of neutral PCommunicators is associated with the higher voting score received in posts.
Thus, our findings show that users tend to upvote more for posts produced by neutral PCommunicators.
This leads to a more influential position for neutral PCommunicators on the comment network. 




\color{black}

\section{Discussion \& Conclusion}

\pb{Summary \& Implications.}
In this paper, we have performed a study on the sharing of partisan news in Reddit during the Russian invasion. We first performed a descriptive analysis of the temporal evolution of partisan news sharing. Our analysis has shown that the Russian invasion stimulated users' discussion in the observed communities. We observed a surge of PSubmissions and comments to these PSubmissions, suggesting more partisan news is spread during the invasion. In addition, commenting behavior becomes more active following breaking events.

We have also characterized partisan news sharing from three aspects -- the appearance of news across multiple subreddits, communities' response to PSubmissions and characteristics of partisan PCommunicators. Our key findings include

\begin{itemize}
    \item \update{An article's partisan bias plays a role in limiting the number of subreddits the partisan news spreads across.}
    News content with a highly 
    partisan leaning is less likely to be shared across multiple subreddits. The findings also aligns with a previous large-scale study on news sharing behavior of Reddit users~\cite{weld2021political}, where the sharing of highly biased news sources tends to be concentrated in a small number of communities.
    
    \item Compared to submissions not containing partisan news, PSubmissions receive more comments. More comments indicate that sharing partisan news can attract more users to engage in the discussion. Moreover,
    posts sharing partisan news seem to attain higher voting scores. These higher voting scores suggest that partisan news can also increase public reception for the posts' content.
    
    \item PCommunicators exhibit different traits to users posting neutral submissions. 
    \update{Compared to neutral PCommunicators, partisan PCommunicators tend to contain more URLs of partisan news in the text.
    In addition, when commenting on partisan news, partisan PCommunicators receive more comments from more distinctive users, on average. Comments by partisan PCommunicators are more likely to be commented on (a higher comment rate). However, it seems that partisan PCommunicators receive lower voting scores on posts. 
    These trends are sufficiently distinct, such that we can use them to identify partisan PCommunicators, with an average accuracy of 61.57\%.}

    \item \update{Neutral PCommunicators have higher PageRank compared to partisan PCommunicators within the comment network. Additionally, a positive correlation exists between PCommunicators' PageRank and the voting scores received by their posts while sharing partisan news. This suggests that, in subreddits related to the Russian invasion, users tend to upvote posts by neutral PCommunicators, resulting in a more influential position for these neutral PCommunicators with the commenting network.}

\end{itemize}

\pb{Limitations.}
\update{As a case study on the ongoing Russian Invasion, this work has limitations. First, our dataset is formed from only six representative Reddit communities related to the invasion. This may reduce our vantage into broader activities.
However, at the time of data collection, these six subreddits are the only ones that are directly related to the warring sides or the events of the Russian Invasion 2022, with a large base of membership (more than 100K members).
Nevertheless, since our investigation solely revolves around the communities associated with the Russian Invasion of 2022, it is less clear how far the observations generalize.}

\update{Second, our study is limited to examining the extent to which users engage with and disseminate partisan news. We do not purport to determine users' inclinations or preferences in terms of supporting news with a specific political bias (left or right). Hence, readers should not assume that every instance of sharing in our study implies an endorsement of a stance. Rather, our focus is on illustrating how partisan news is propagated during the Russian Invasion 2022, and how the associated communities react to it.}

\pb{Future work.}\update{The findings presented in this study represent only a preliminary step towards the understanding of the sharing of partisan news and the interactions between news propagators and consumers. Specifically, our investigation is limited to metrics such as the number of comments and voting scores.
In our future work, we plan to accompany this with a content analysis of users' discussions related to partisan news. 
This could shed more nuanced light on how users interact with partisan submissions.}

\pb{Ethics consideration.}We use public data from the Pushshift API~\cite{baumgartner2020pushshift}, which is widely used for research based on the Reddit platform. Our work follows the Reddit's privacy terms.

\bibliography{aaai22}
\end{document}